\documentclass[twocolumn,tighten]{aastex62}
\usepackage{natbib}

\usepackage[hang,raggedright,tight]{subfigure}
\usepackage{longtable}
\usepackage[figuresright]{rotating}
\usepackage{float}
\usepackage[toc,page]{appendix}

\usepackage{gensymb}
\usepackage{threeparttable}
\usepackage{diagbox}
\usepackage{pbox}
\usepackage{multirow}  
 
\shorttitle{decay-rate }
\shortauthors{Dong et al.}

\begin{document}
\title{The direct measurement of gravitational potential decay rate at cosmological scales II -- Improved dark energy constraint from $z\le1.4$}
\author[0000-0003-0296-0841]{Fuyu Dong}
\altaffiliation{dfy@ynu.edu.cn}
\affiliation{South-Western Institute for Astronomy Research, Yunnan University, Kunming 650500, China}
\author{Pengjie Zhang}
\altaffiliation{zhangpj@sjtu.edu.cn}
\affiliation{Department of Astronomy, School of Physics and Astronomy, Shanghai Jiao Tong University, Shanghai, 200240, China} 
\affiliation{Division of Astronomy and Astrophysics, Tsung-Dao Lee Institute, Shanghai Jiao Tong University, Shanghai, 200240, China}
\affiliation{Key Laboratory for Particle Astrophysics and Cosmology
(MOE)/Shanghai Key Laboratory for Particle Physics and Cosmology,
China} 
\author{Haojie xu}
\affiliation{Shanghai Astronomical Observatory, Chinese Academy of Sciences, Shanghai 200030, China} 
\author{Jian Qin}
\affiliation{Department of Astronomy, School of Physics and Astronomy, Shanghai Jiao Tong University, Shanghai, 200240, China} 

\begin{abstract}

The gravitational potential decay rate (DR) is caused by the cosmic acceleration of the universe, providing a direct probe into the existence of dark energy (DE). We present measurements of DR and explore its implications for DE models using the Data Release 9 galaxy catalog of DESI imaging surveys and the Planck cosmic microwave background maps. Our analysis includes six redshift bins within the range of $0.2\le z<1.4$ and achieves a total significance of 3.1$\sigma$,  extending the DR measurements to a much higher redshift comparing to Dong et al. (2022), which focused on $0.2\le z<0.8$.  Other improvements involve addressing potential systematics in the DR-related measurements of correlation functions, including imaging systematics and magnification bias. We explore the constraining power of DR both the $w$CDM model and the $w_0w_a$CDM model. We find that, the addition of DR can
significantly improves DE constraints, over Sloan Digital Sky Survey baryon acoustic oscillation (BAO) data alone or PantheonPlus supernovae (SNe) compilation alone, although it shows only a modest improvement for DESI BAO. In the $w$CDM model, all three probes-DR, DESI BAO and SNe-favor $w=-1$. For the $w_0w_a$CDM, while DESI BAO prefers $w_0>-1$ and $w_a<0$, SNe Ia and DR data constrain $w_0=-0.94^{+0.11}_{-0.13}$ and $w_a=-0.22^{+0.57}_{-0.97}$. Namely SNe Ia and DR data has no preference on dynamical dark energy over  $\Lambda$. 
\end{abstract}

\section{Introduction}
Dark Energy (DE) accounts for approximately 70\% of the energy density of the universe and drives the acceleration of cosmic expansion; in turn, this affects the redshift-distance relation and growth of structure. While DE has played a significant role in recent epochs, its effect at high redshift is expected to be minimal. Consequently, the redshift range that is most effective for probing dark energy is expected to be less than two\citep{2001PhRvD..64l3527H}. There are several methods that show particular promise for investigating DE within this redshift range. For example, distance measurements from Type Ia supernovae (SN Ia)\citep{1998AJ....116.1009R,1999ApJ...517..565P,2009ApJ...699..539R,Freedman_2012}, anisotropies in the cosmic microwave background (CMB) radiation\citep{Balbi_2000,2000Natur.404..955D,2007ApOpt..46.3444F,2011PASP..123..568C,2013ApJS..208...20B}, and the baryon acoustic oscillation (BAO) feature in galaxy clustering\citep{2005MNRAS.362..505C,2005ApJ...633..560E,2012MNRAS.427.3435A} have largely confirmed the accelerated expansion of the universe. Among the various probes of DE, the integrated Sachs–Wolfe (ISW) effect\citep{1967ApJ...147...73S} provides a powerful evidence for an accelerated expansion in a flat universe. 

The ISW effect refers to the distortion of the CMB temperature caused by the time evolution of the gravitational potential sourced by large-scale structures (LSS) along the line of sight. The method for observing this secondary anisotropy of the CMB is through cross correlating the CMB with large-scale structure (LSS)\citep{1996PhRvL..76..575C,2000ApJ...538...57S,2003ApJ...597L..89F,2004Natur.427...45B,2004PhRvD..70h3536A,2004PhRvD..69h3524A,2004MNRAS.350L..37F,2004ApJ...608...10N,2005NewAR..49...75B,2005PhRvD..72d3525P,2005PhRvD..71l3521C,2006MNRAS.365..891V,2006astro.ph..2398M,2007MNRAS.381.1347C,2007MNRAS.377.1085R,2008MNRAS.386.2161R,2010A&A...520A.101H,2010MNRAS.404..532M,2012MNRAS.427.3044S,2012MNRAS.426.2581G,2014A&A...571A..19P,Shajib_2016,2016A&A...594A..21P,PhysRevD.97.103514,2019MNRAS.484.5267K,2021MNRAS.500.3838D,2022MNRAS.tmp.2016B}. However, such a measurement suffers from the degeneracy between $\dot{\phi}$ and galaxy bias/matter clustering, limiting its usefulness in constraining dark energy. In order to isolate the decay rate (DR) of gravitational potential, \cite{2006ApJ...647...55Z} proposed a novel method  by establishing a proportionality relation between the ISW-LSS cross-power spectrum $C_{Ig}$ and lensing-LSS cross-power spectrum $C_{\phi g}$. The ratio of these two cross-correlations measures $\dot{\phi}/\phi$, up to a prefactor depending on the geometry of the universe but free of galaxy bias and matter clustering:
\begin{equation}
\label{eq:dr-cl}
C(\ell)_{Ig}\simeq {\rm DR}(z_m)C(\ell)_{\phi g},
\end{equation}
where $z_m$ is the average redshift over the redshift range $[z_m-\Delta z/2,z_m+\Delta z/2]$. The coefﬁcient DR is the decay rate that to be measured:
\begin{equation}
{\rm DR}(z)=\left(-\frac{d\,\rm{ln}D_\phi}{d\,\rm{ln}a}\right)\left(\frac{aH(z)/c}{W_L(z)}\right)
\end{equation}


DR deﬁned above differs from the desired decay rate $d{\rm ln}D_\phi/d{\rm ln}a$ by extra factors in the last parenthesis, which has no dependence on the value of $H_0$ and thus avoid uncertainties in $H_0$. It also has the benefit of being insensitive to the sampling variance since the lensing-LSS and ISW-LSS measurements use the same selection function over the same volume. We achieved the first direct detection of DR with a total signiﬁcance of $\sim2.7\sigma$ at three redshifts $z_m=0.3/0.5/0.7$ in \cite{2022ApJ...938...72D} based on the DESI DR8 photometric galaxy catalog, and have successfully used it to constrain the $w$CDM model. Due to the intrinsic dependence of decay rate on $\dot{\phi}/\phi$, its sensitivity to $w$ is a factor of 3/4/5 higher at $z=0.3/0.5/0.7$ than that of $H(z)$. Consequently, despite its relatively low signal-to-noise ratio, the addition of DR was found to signiﬁcantly improve DE constraints, over Sloan Digital Sky Survey (SDSS) baryon acoustic oscillation (BAO) data alone or Pantheon supernovae (SNe) compilation alone. 

Recently, \cite{2024arXiv240403002D} has reported the BAO measurements with the first year of data from DESI. The region of the sky and the range of redshifts observed by DESI partially overlap with those covered by the previous generation of BOSS \citep{2013AJ....145...10D} and eBOSS \citep{2016AJ....151...44D} survey programs from SDSS.
Approximately $\sim 70\%$ of the DESI DR1 footprint was covered by BOSS; while about $\sim 65\%$ of the BOSS footprint has been covered by DESI DR1.  Although the input catalogs used in DESI DR1 BAO analyses and SDSS BAO analyses share a fraction of common objects, discrepancies have been observed, the cause of which remains clear\citep{2024arXiv240403002D}. 
Therefore, it would be intriguing to analyze the DR results alongside the DESI DR1 BAO results to determine whether the cosmological constraints are influenced.


DR measurements from higher redshift must be included, nevertheless, for the reasons listed below. First, for flat $w$CDM model, the sensitivity of DR to $w$ monotonically increases with redshift. At $z=1.4$, it approaches 12 times the sensitivity of $H$, and 2.1 times that of DR itself at $z=0.8$. For this reason, DR from $z\ge0.8$ redshift will further improve the DE constraint. Second, even a null DR detection presented at high $z$ is helpful to distinguish distinct dark energy models.  Third, more independent volumes are available for higher $z$, meaning lower statistical error. Therefore, it is essential to investigate whether the DR measurements from high redshift can impact the DE constraints from other probes.

In this work we will measure DR to a redshift of 1.4 using DR9 galaxy catalog of DESI imaging surveys. At high redshifts, the magnification bias effect becomes significant. This bias arises from the presence of mass along the line of sight (LOS) between redshift $z$ and $0$, resulting in an additional term in the intrinsic galaxy density. Additionally, observational conditions can introduce spurious fluctuations in the observed density of photometric galaxy samples\citep{Scranton_2002,Myers_2006,10.1111/j.1365-2966.2011.19351.x, Ho_2012,10.1093/mnras/stv2103,10.1111/j.1365-2966.2011.19351.x}. Consequently,  both $C_{Ig}$ and $C_{\phi g}$ are biased due to the correlation between the background CMB lensing/ISW signal and the extra foreground galaxy density. Therefore, we will incorporate treatments of these two potential systematics for the DR measurements.

This paper is organized as follows. Section \ref{sec:data} introduces the data sets used in our analysis. Section 3 presents the measurements of DR at six redshifts, as well as its constraints on the flat $w$CDM model and a flat $w_0w_a$ DE model. We discuss and conclude in Section 4. 

\section{Data and Systematics Mitigation}
\label{sec:data}

\begin{figure}
    \centering
     \subfigure{
     \includegraphics[width=1\linewidth, clip]{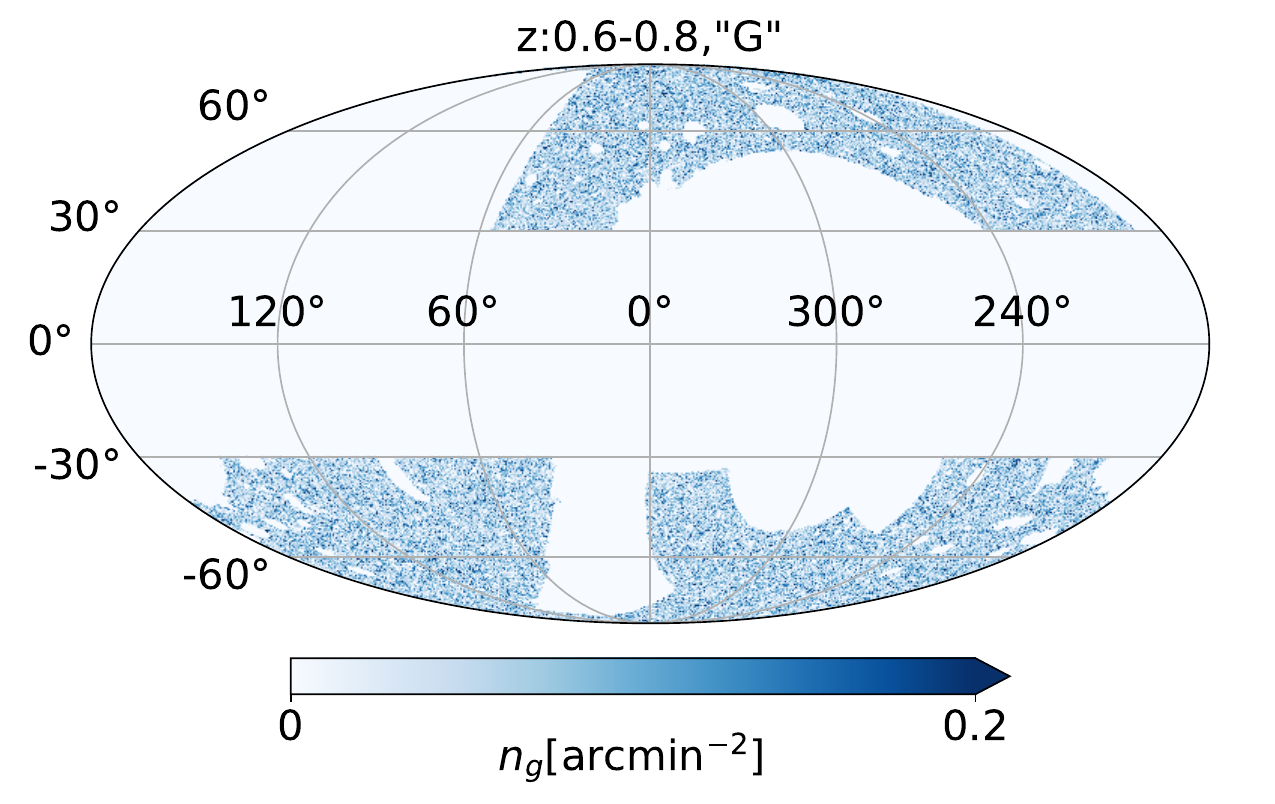}}
   \caption{Source distribution in the DR9 catalog of the DECaLS+DES imaging surveys adopted in our measurement. Both CMB mask and galaxy survey mask have been adopted, along with additional masks near the Large Magellanic Cloud, the disconnected areas as well as the boarders surrounding DES. The depth of color represents the number of galaxies per $\rm{arcmin^2}$.} 
    \label{fig:gmap}
\end{figure}

\begin{figure*}
    \centering
     \subfigure{
     \includegraphics[width=1\linewidth, clip]{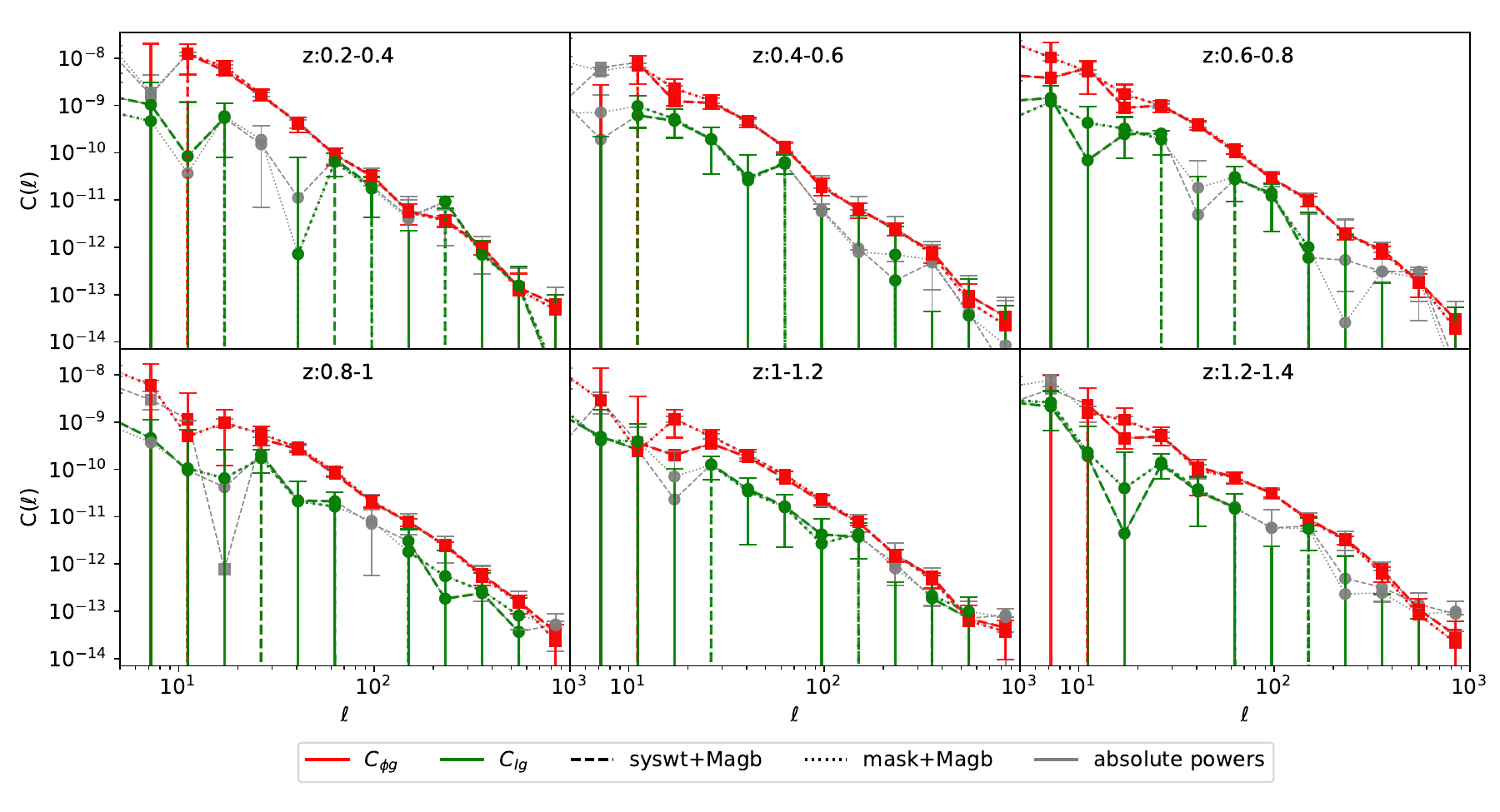}}
   \caption{The cross-correlation between CMB lensing (or ISW) and galaxies measured for six redshift bins: $z_m$ = 0.3, 0.5, 0.7, 0.9, 1.1 and 1.3. The red and green color show results of $C_{\phi g}$ and $C_{Ig}$, respectively. There are two line styles shown for each color. The dashed line shows the signal measured by adopting the imaging weight, while the dotted line shows the signal without adopting any weight but the using same angular mask. For comparison, we have inverted the negative powers and displayed them in gray. All the results have been calibrated for the magnification bias.} 
    \label{fig:cl_iswphiallz}
\end{figure*}

\subsection{CMB Temperature Map and Lensing data}
We use the Planck SMICA temperature product \citep{2016A&A...594A..11P} to measure the ISW-galaxy cross-correlation. For the lensing potential-galaxy cross-correlation measurement, we use the Planck lensing product\citep{2020A&A...641A...8P}. For both products, the thermal Sunyaev-Zeldovich (tSZ) effect has been deprojected. The Planck analysis mask is adopted to remove pixels contaminated by Galactic dust or known points sources. 

Notice that to avoid systematic biases to the lensing data introduced by the mask effect,  aliasing effect and reconstruction noise, we apply a tophat cut in multipole space for generating the lensing potential for which we refer readers to the Appendix B of \cite{2022ApJ...938...72D}. The same filter is applied to the CMB temperature to eliminate its effect on the DR measurement. We then downgrade the above CMB products to a lower resolution with Nside = 512 to match the resolution of the galaxy density map.

\subsection{Galaxy Catalogue}
\label{sec:galaxy}
We perform the analysis based on the public Data Release 9 galaxy catalog with photometric redshifts obtained from the DESI Legacy Imaging Surveys. The optical imaging data (g,r,z bands) are provided by three projects, the Beijing-Arizona Sky Survey (BASS), the Mayall z-band Legacy Survey (MzLS), and the DECam Legacy Survey (DECaLS), while the Wide-ﬁeld Infrared Survey Explorer (WISE) satellite provides the infrared data. In combination with the  Dark Energy Surey (DES), these four surveys covers ~20000 $\rm{deg}^2$ sky area in total.

The DR measurements in this work are performed on six evenly spaced tomographic redshift bins at $0.2\le z<1.4$. To ensure good uniformity in the distribution of the galaxy sample within the survey area, we will select galaxies from the DESI + DES region, with z-band absolute magnitudes $M_z<-22$\footnote{We adopt the fit function to the average $K_z^{0.5}$ values in \cite{2021ApJ...909..143Y} to do an overall K-correction. $K_z^{0.5}$ represents the K-correction in z band to redshift $z\sim0.5$.}.  Furthermore, we add an additional mask to maps to remove regions within $|b|<30.0^\circ$ (where $b$ is the Galactic latitude), considering that both the reconstructed Planck CMB lensing maps and galaxy data may suffer from contamination from high stellar density regions. We show an example of the galaxy distribution in Fig.\ref{fig:gmap}, generated with $0.6\le z<0.8$. A combined mask from both galaxy catalog\footnote{The survey mask is generated from the available random catalogs provided by the DR9 website, which contain the number of observation in the g, r, z bands and a general purpose artifact ﬂag MASKBITS, according to the coordinates drawn from the observed distribution. We refer readers to \cite{2021ApJ...923..153D} for the detailed procedure.} and the Planck survey is applied to the galaxy density ﬁeld. 

\subsection{Systematics mitigation}
\label{sec:systematic}
The large scale clustering of photometric samples are potentially biased in observation, due to spurious fluctuations in galaxy density induced by observation conditions, such as Galactic extinction, stellar contamination and so on. In this work, we employ a machine learning method to mitigate the spurious correlations and improve the reliability of clustering measurements\citep{2023MNRAS.520..161X,2022MNRAS.509.3904C}. During this process, we further mask out the following regions here, where the right ascension (ra) and declination (dec) fall within [${\rm ra_1, ra_2}$, $\rm{dec_1, dec_2}$]: $[52^\circ, \; 120^\circ, \; -90^\circ, \; -50^\circ]$, $[120^\circ, \; 150^\circ, \;-45^\circ, \; -10^\circ]$, $[150^\circ, \; 180^\circ, \; -45^\circ, \; -15^\circ]$, \newline
$[210^\circ, \; 240^\circ, \; -20^\circ, \; -12^\circ]$, 
$[180^\circ, \; 240^\circ, \; -90^\circ,-18.5^\circ]$, \newline
$[120^\circ, \; 180^\circ, \;-90^\circ, \; -17.4^\circ]$\footnote{The first coordinate range is used to mask out the LMC area, while the second to fourth ranges mask out the disconnected areas in the NGC South. The last two ranges are intended to mask the borders of the footprint surrounding DES.}. This includes areas near the Large Magellanic Clouds, as well as a DES-related mask. 

The additional angular masks here and the selection criterion in Section \ref{sec:galaxy} result in a total number of $\sim 1.4\times10^7$ galaxies distributed across six evenly spaced redshift bins for $0.2\le<z<1.4$, with the final sky coverage being 11183 ${\rm deg^2}$. For each redshift, we apply the Random Forest mitigation (RF) technique of \citep{2022MNRAS.509.3904C} to our galaxy samples and derive a correction weight to each galaxy. The imaging maps used in this work were provided by E. Chaussidon, which are generated from the DESITARGET\footnote{\url{https://github.com/desihub/desit arget}} package with HEALPix $N_{\rm{side}}=256$. 

The following photometric properties are selected as our imaging features: stellar density, Galactic extinction, sky brightness, exposure time, PSF size and PSF depth(g/r/z/W1/W2). We bin galaxies into density maps with the same resolution as the input imaging maps. We also adopt the same RF hyperparameters as in \cite{2022MNRAS.509.3904C}, for which we refer readers to \cite{2023MNRAS.520..161X} for further details. Finally, we obtain a weight factor for each valid pixel. The imaging systematics will then be reduced by weighting each galaxy according to its respective pixel weight. 
In Fig.\ref{fig:cl_iswphiallz}, we show the tomographic cross-correlation measurements of galaxy-ISW and galaxy-lensing at six redshift bins. As shown, (1) both the $C_\ell^{g\phi}$ and $C_\ell^{g\rm{I}}$ are statistically suppressed at small $\ell$; (2) $C_\ell^{g\phi}$ is enhanced at large $\ell$, while $C_\ell^{g\rm{I}}$ remains suppressed at large $\ell$ in most cases; and (3) the error bar slightly increases after adopting the imaging weight.

In addition to the imaging systematics, the spatial distribution of galaxies at high redshift will also suffer from serious lensing magnification bias, as weak lensing affects the observed galaxy number over-density: 
\begin{equation}
\delta_g^L=\delta_g+q\kappa,
\end{equation}
where $\delta_g=n_g/\overline{n}-1$ denote the intrinsic projected galaxy over-density and $\kappa$ the lensing convergence. For a flux-limited sample, $q=5s-2$ and $s$ is the magnitude slope of the cumulative number counts at the faint end cut-off:
\begin{equation}
s=\frac{d\rm{log_{10}}n_g(<{\rm m})}{d{\rm m}}.
\end{equation}

In this paper, we find that $s\sim 1$ for our galaxy samples at all redshifts. Therefore, throughout the remainder of the paper we adopt $q=3$ and the Planck cosmology to theoretically estimate the impact of magnification bias on the measured cross-powers $C_{\phi g}$ and $C_{Ig}$ from observation. We cross-check the value of $q$ using the galaxy-shear cross-correlation in Sec.\ref{sec:conclusion} and find them to be consistent.

\section{DR measurements and improved dark energy constraints}
\label{sec:result}
\begin{figure*}
    \centering
     \subfigure{
     \includegraphics[width=1\linewidth, clip]{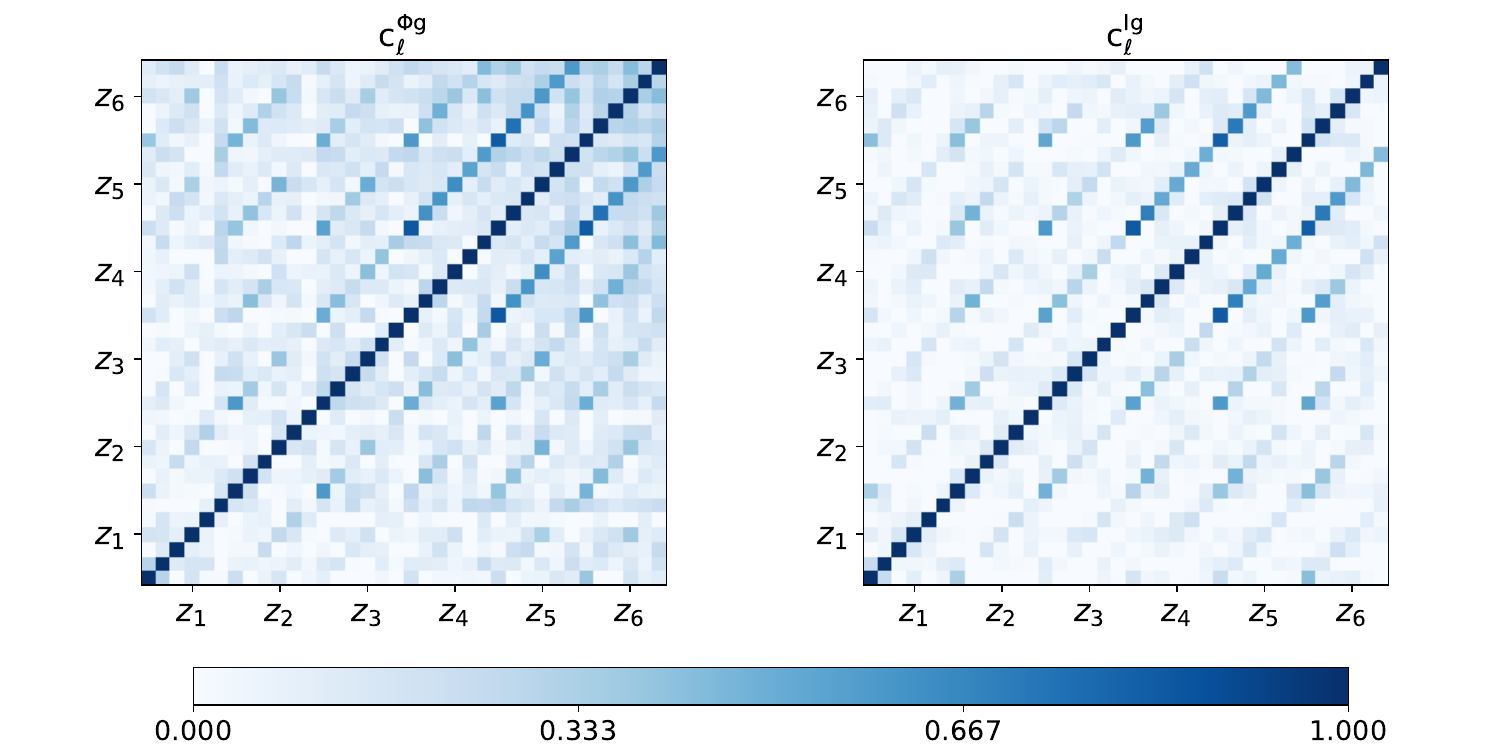}}
   \caption{The covariance matrix for $C_{\phi g}$ and $C_{Ig}$, for which the cross-powers are calculated by adopting the imaging weight.} 
    \label{fig:ncov_6z}
\end{figure*}

\begin{figure}
    \centering
     \subfigure{
     \includegraphics[width=1\linewidth, clip]{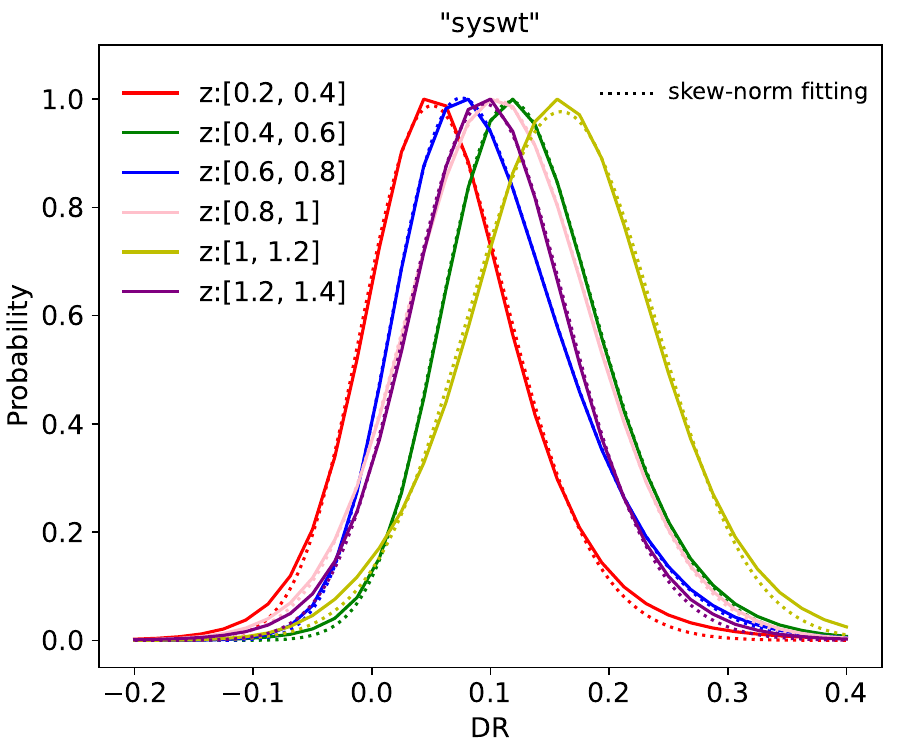}}
   \caption{The probability distribution function of ${\rm DR}(z)$ obtained from a Bayesian analysis (solid line). It is well described by a skew-normal function (dashed line). The PDF is normalized with its peak amplitude. }
    \label{fig:DR-PDF}
\end{figure}

\begin{table*}[]
\begin{minipage}{180mm}
\centering
\caption{The galaxy samples selected from DR9 south DESI imaging survey (DECaLS+DES) and the measured gravitational potential decay rate DR. We adopt a z-band abs magnitude cut on the faint end with -22. For the measurement of DR, we choose to use $C_{Ig}$ and $C_\phi g$ distributed within ($\ell_{min}$, $\ell_{max})\sim(9,117)$ and angular bins equally spaced in $ln\ell$.}
\begin{tabular}{cc|cc|cc}
\hline
 &  & \multicolumn{2}{c}{Magnification-bias calibrated} & \multicolumn{2}{c}{w/o Magnification-bias calibration}   \\
\hline
Galaxy redshift & $N_{\rm g}$ &DR-"syswt" & DR-"mask"  &DR-"syswt" &DR-"mask" \\
\hline
$0.2\le z^P<0.4\,(z_m\simeq0.31)$ & $0.65\times 10^6$ & $\mathbf{0.047^{+0.066}_{-0.057}}$ & $0.029^{+0.060}_{-0.051}$ & $0.050^{+0.066}_{-0.057}$ & $0.032^{+0.057}_{-0.051}$\\
$0.4\le z^P<0.6\,(z_m\simeq0.51)$ & $1.32\times 10^6$ & $\mathbf{0.114^{+0.072}_{-0.063}}$ & $0.126^{+0.069}_{-0.057}$ & $0.117^{+0.072}_{-0.057}$ & $0.126^{+0.066}_{-0.054}$\\
$0.6\le z^P<0.8\,(z_m\simeq0.70)$ & $2.47\times 10^6$ & $\mathbf{0.068^{+0.087}_{-0.060}}$ & $0.083^{+0.069}_{-0.083}$ & $0.074^{+0.078}_{-0.057}$ & $0.086^{+0.054}_{-0.063}$\\
$0.8\le z^P<1.0\,(z_m\simeq0.91)$ & $4.85\times 10^6$ & $\mathbf{0.105^{+0.087}_{-0.084}}$ & $0.068^{+0.069}_{-0.069}$ & $0.105^{+0.078}_{-0.075}$ & $0.071^{+0.063}_{-0.066}$\\
$1.0\le z^P<1.2\,(z_m\simeq1.09)$ & $3.87\times 10^6$ & $\mathbf{0.165^{+0.099}_{-0.103}}$ & $0.080^{+0.078}_{-0.087}$ & $0.159^{+0.081}_{-0.078}$ & $0.089^{+0.072}_{-0.066}$\\
$1.2\le z^P<1.4\,(z_m\simeq1.28)$ & $1.33\times 10^6$ & $\mathbf{0.086^{+0.081}_{-0.081}}$ & $0.071^{+0.084}_{-0.075}$ & $0.095^{+0.069}_{-0.066}$ & $0.080^{+0.069}_{-0.063}$\\
\hline
\hline
\end{tabular}
\label{tab:DR}
\begin{tablenotes}
      \small
      \item Note. --- Columns for $N_g$, DR, $\langle DR\rangle$ and $\sigma(DR)$ refers to the number of galaxies being used, the best-estimate value of DR, the average value and the scatter with respect to $\langle DR\rangle$. We show the baseline results obtained by calibrating the magnification bias with a fixed cosmology of $\Omega_m=0.315$ and $w=-1$.
    \end{tablenotes}
\end{minipage}
\end{table*}

\begin{table}[]
\centering
\caption{Similar to Table.\ref{tab:DR}, but with DR measured independently from $z_{1,2,3}$ and $z_{4,5,6}$. We have also calibrated the magnification bias here.}
\begin{tabular}{c|c}
\hline
Galaxy redshift  & DR-"syswt" \\
\hline
$0.2<z^P<0.4$  & $0.047^{+0.066}_{-0.069}$\\
$0.4<z^P<0.6$  & $0.138^{+0.081}_{-0.063}$ \\
$0.6<z^P<0.8$  & $0.117^{+0.084}_{-0.069}$ \\
\hline 
$0.8<z^P<1.0$  & $0.098^{+0.090}_{-0.084}$ \\
$1.0<z^P<1.2$  & $0.156^{+0.094}_{-0.090}$ \\
$1.2<z^P<1.4$  & $0.065^{+0.084}_{-0.072}$ \\
\hline
\end{tabular}
\label{tab:DR-3}
\end{table}

\begin{figure*}
    \centering
     \subfigure{
     \includegraphics[width=0.45\linewidth, clip]{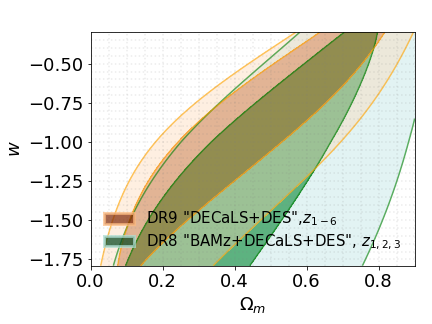}} 
     \subfigure{
     \includegraphics[width=0.45\linewidth, clip]{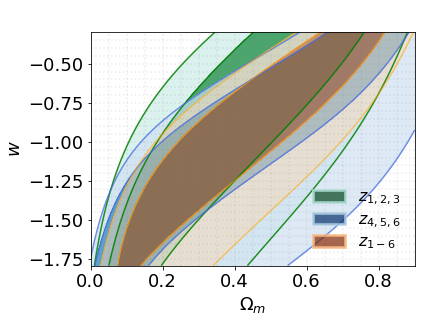}}
   \caption{Constraints on the flat $w$CDM model from DR alone. In the left panel, we show the result by using all the DR measurements from six redshifts (i.e., ${\rm DR}_{6z}$) in the red contour, based on galaxy samples from the DESI DR9 "DECaLS+DES" survey areas. For comparison, we also plot the result reported in \cite{2022ApJ...938...72D}  using the ${\rm DR}_{1,2,3}$ (i.e., DR at the first three redshifts) measurements from the DESI DR8 "BAMz+DECaLS+DES" survey areas. The right panel shows the DE constraints in this work using DR measurements across different redshift choices. The addition of $z_{4,5,6}$ is useful to improve the DE constraints.}
\label{fig:DR9-DR8}
\end{figure*}

\begin{figure*}
    \centering
     \subfigure{
     \includegraphics[width=0.43\linewidth, clip]{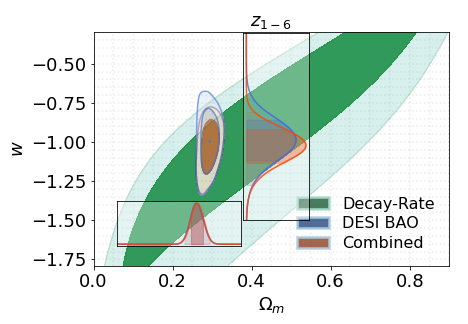}}
     \subfigure{
     \includegraphics[width=0.43\linewidth, clip]{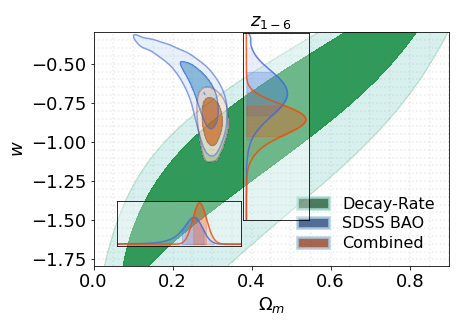}}
   \caption{Constraints on the flat $w$CDM model, from BAO/DR/BAO+DR. The left panel is obtained with the DESI BAO data, while the right panel is obtained with the SDSS BAO data. In both panels, DR shows a different $\Omega_m$-$w$ degeneracy direction compared to BAO data. When compared to SDSS BAO alone, the inclusion of DR significantly improves the constraints; however it only provides a modest improvement over the DESI BAO.}
    \label{fig:6z-syswt-wcdmcontour-BAO}
\end{figure*}
\begin{figure}
    \centering
      \subfigure{
     \includegraphics[width=1\linewidth, clip]{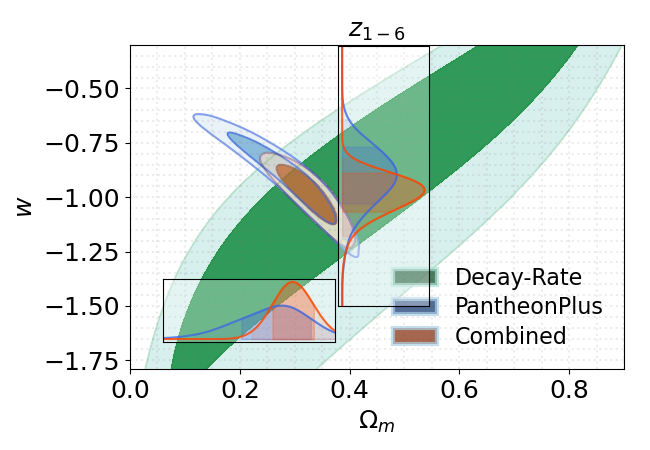}}
   \caption{Similar to Fig.\ref{fig:6z-syswt-wcdmcontour-BAO}, but with constraints from SNe/DR/SNe+DR.}
    \label{fig:6z-syswt-wcdmcontour-SN}
\end{figure}

\subsection{Measurement of DR}
We use a Bayesian analysis \citep{2022ApJ...938...72D,2023ApJS..267...21S} for estimating DR, instead of directly taking the ratio of $C_{Ig}$ and $C_{\phi g}$. This likelihood method directly evaluate the PDF $P(DR)$ using the exact analytical expression and involves no multiparameter ﬁtting. Considering that $C_{Ig}$ mainly arises from large angular scales, we choose to use the cross-powers within the range of $(\ell_{min},\ell_{max})\sim(9,117)$ for the DR measurements. 

In Fig.\ref{fig:ncov_6z} we show the full covariance matrices respectively for $C_{Ig}$ and $C_{\phi g}$, which includes both the auto-correlations and cross-correlations between different redshift bins. Strong cross-correlations are found between redshifts [$z_4$,$z_5$], as well as redshifts [$z_5$,$z_6$], possibly caused by a mixture of galaxies due to the lower accuracy in the photo-z estimation for higher redshifts. Besides, nonzero cross-correlations are found at large scales (small $\ell$) among different redshifts, which are possibly caused by the same angular mask, and the wide ISW and CMB lensing kernels. 

Considering the non-negligible cross-correlations, we measure DR($z$) at six redshifts simultaneously using the full covariance matrix. The results are shown in Fig.\ref{fig:DR-PDF} and Table.\ref{tab:DR}. The total significance of the DR meausrements is $\sim 3.1\sigma$.
We then fit $P(DR,z_i)$ with a skew-norm form in order to obtain a continuous probability distribution (PDF) function of DR($z$).

\subsection{Constraint on a Flat $w$CDM Model}
\subsubsection{Constraint from DR}
In this section, we restrict the analysis to the flat $w$CDM cosmology for the DE constraint, in which DR only relies on two cosmological parameters: $\theta=\{\Omega_m,w\}$. We estimate the posterior distribution of $\theta$ with DR from redshifts $z_{n_1}\le z\le  z_{n_2}$ as:
\begin{equation}
P(\theta|{\rm DR})\propto\prod\limits_{i=n_1}^{n_2}P_i({\rm DR}|\theta,z_i)\pi(\theta),
\end{equation}
where $1\le n_1<n_2\le6$. We employ uniform priors on $\Omega_m$ and $w$ within the ranges of [0,0.9] and [-1.8,-0.3], i.e., $\pi(\theta)=const$.  For a given DR, $P_i({\rm DR}|\theta,z_i)$ can be obtained from the PDF of ${\rm DR}(z_i)$ (Fig.\ref{fig:DR-PDF}).  Fig.\ref{fig:DR9-DR8} shows the DE constraints from a joint analysis using DR measurements across different redshift choices. We consider three cases for the joint analysis: 1) using DR estimated from all six redshift slices $z_{1-6}$, labeled as ${\rm DR}_{6z}$; 2) using DR from the first three redshift slices $z_{1,2,3}$, labelled as ${\rm DR}_{f3z}$; and 3) using DR of the last three redshift slices $z_{4,5,6}$, labelled as ${\rm DR}_{l3z}$. In the following, we will use ${\rm DR}_{6z}$ as our baseline result to analyze the constraining power of DR on DE.  

We first revisit DR measured from $z_{1,2,3}$, taking into account the changes in the galaxy samples compared to \cite{2022ApJ...938...72D}. In this analysis, we measure $\rm{DR}_{f3z}$ simultaneously using the corresponding covariance matrices $cov(z_i,z_j)$, where $1\le i,j\le3$.
The results are presented in Table.\ref{tab:DR-3}. We find that the total significance of ${\rm DR}_{f3z}$ is approximately $\sim2.56\sigma$, which is slightly lower than the previously reported $2.67\sigma$ (joint analysis). This decrease in S/N of ${\rm DR}_{f3z}$ is due to the exclusion of the BAMz survey area in this work (BAMz refers to "BASS+MzLS"). Consequently, the size of survey areas used for measuring DR is reduced by 27\%, leading to an expected S/N of $\sim 2.3\sigma$. This value is lower than the 2.56$\sigma$ detection reported here, which can be attributed to the systematic calibrations applied to the DR measurements in this study, as well as to the cosmic variance introduced by the exclusion of areas corresponding to the BAMz survey from the CMB temperature map, since the CMB is the main source of noise in ISW measurements.

Next, we measure DR from $z_{4,5,6}$ and find the total significance to be $2.2\sigma$. Therefore, a total significance of ${\rm DR}_{6z}$ is expected to be $3.37\sigma$ (calculated as $\sqrt{(S/N)^2_{f3z}+(S/N)^2_{l3z}}$). However, we find that the S/N of ${\rm DR}_{6z}$, measured using the full covariance matrices, is $\sim3.1\sigma$, as the cross-correlations between low redshifts and high redshifts will further reduce $\rm{(S/N)}_{6z}$. Although ${\rm DR}_{6z}$ shows a modest improvement over the previous $2.67\sigma$ detection of ${\rm DR}_{f3z}$, it does yield an improved constraint on DE(left panel of Fig.\ref{fig:DR9-DR8}).  One factor that leads to the DE constraints being stronger than what would be implied by the measured S/N is that the sensitivities of DR to $w$ increase rapidly with redshift, which is a factor of 2.4 at $z=1.4$ over $z=0.8$. Therefore, the $\Omega_m-w$ constraint from ${\rm DR}_{l3z}$ is partly complementary to that of ${\rm DR}_{f3z}$, as is shown in the right panel of Fig.\ref{fig:DR9-DR8}. The inclusion of $\rm{DR}_{l3z}$ measurements thus improves the overall constraining power on $w$.


\begin{table*}
\centering
\caption{Marginalized values and 68$\%$ confidence limits of ($\Omega_m$, w) estimated from DR when in combination with different probes and with different choices of redshift. \label{Table3}}
\begin{minipage}{180mm}
\centering
\begin{tabular}{ccc|cc|cc}
\hline
\hline
&Probes & Magb &$\Omega_m$ (best),  & $\langle\Omega_m\rangle$,  &$w$ (best), & $\langle w \rangle$ \\
\hline
     -      &DESI BAO      &-       & $0.290_{-0.013}^{+0.016}$ &$0.292\pm0.015$ &$-0.981_{-0.135}^{+0.127}$ &$-0.991\pm0.138$ \\
"syswt"(6z) &BAO+DR &\emph{Base.} & $0.292_{-0.014}^{+0.014}$ &$0.293\pm0.015$ &$-1.019_{-0.120}^{+0.112}$ &$-1.031\pm0.119$ \\
"mask" (6z) &BAO+DR &\emph{Base.} & $0.292_{-0.014}^{+0.014}$ &$0.293\pm0.015$ &$-1.064_{-0.112}^{+0.105}$ &$-1.076\pm0.116$ \\
"syswt"(6z) &BAO+DR &\emph{Vary.} & $0.292_{-0.014}^{+0.013}$ &$0.293\pm0.015$ &$-0.999_{-0.120}^{+0.097}$ &$-1.010\pm0.115$ \\
"syswt"(f3z)&BAO+DR &\emph{Base.} & $0.290_{-0.014}^{+0.013}$ &$0.292\pm0.015$ &$-0.989_{-0.127}^{+0.127}$,&$-0.998\pm0.130$\\
"syswt"(l3z)&BAO+DR &\emph{Base.} & $0.290_{-0.014}^{+0.013}$ &$0.292\pm0.015$ &$-1.004_{-0.120}^{+0.112}$,&$-1.021\pm0.121$\\
"syswt"(fl3z)&BAO+DR &\emph{Base.} & $0.292_{-0.014}^{+0.013}$ &$0.292\pm0.015$ &$-1.019_{-0.120}^{+0.112}$,&$-1.033\pm0.119$\\

\hline
     -      &SDSS BAO      &-       & $0.285_{-0.030}^{+0.024}$ &$0.272\pm0.035$ &$-0.682_{-0.14}^{+0.13}$   &$-0.69\pm0.144$ \\
"syswt"(6z) &BAO+DR &\emph{Base.} & $0.299_{-0.016}^{+0.016}$ &$0.299\pm0.017$ &$-0.854_{-0.105}^{+0.090}$ &$-0.871\pm0.104$ \\
"mask" (6z) &BAO+DR &\emph{Base.} & $0.301_{-0.014}^{+0.016}$ &$0.302\pm0.017$ &$-0.906_{-0.112}^{+0.089}$ &$-0.927\pm0.106$ \\
"syswt"(f3z)&BAO+DR &\emph{Base.} & $0.290_{-0.018}^{+0.020}$ &$0.289\pm0.021$ &$-0.757_{-0.120}^{+0.120}$,&$-0.769\pm0.123$ \\
"syswt"(l3z)&BAO+DR &\emph{Base.} & $0.296_{-0.016}^{+0.016}$ &$0.297\pm0.017$ &$-0.832_{-0.105}^{+0.097}$,&$-0.845\pm0.106$ \\
\hline
    -       &PantheonPlus       &-       & $0.307_{-0.070}^{+0.051}$ &$0.287\pm0.064$ &$-0.892_{-0.135}^{+0.127}$,&$-0.908\pm0.138$ \\
"syswt"(6z) &SN+DR  &\emph{Base.} & $0.323_{-0.034}^{+0.038}$ &$0.326\pm0.037$ &$-0.959_{-0.097}^{+0.075}$,&$-0.985\pm0.092$ \\
"mask" (6z) &SN+DR  &\emph{Base.} & $0.341_{-0.032}^{+0.034}$ &$0.343\pm0.034$ &$-1.011_{-0.097}^{+0.082}$,&$-1.031\pm0.093$ \\
"syswt"(f3z)&SN+DR  &\emph{Base.} & $0.312_{-0.049}^{+0.045}$ &$0.306\pm0.047$ &$-0.921_{-0.112}^{+0.097}$,&$-0.942\pm0.108$ \\
"syswt"(l3z)&SN+DR  &\emph{Base.} & $0.305_{-0.054}^{+0.045}$ &$0.297\pm0.049$ &$-0.899_{-0.112}^{+0.097}$,&$-0.922\pm0.110$ \\
\hline
\hline
\end{tabular}
\begin{tablenotes}
      \small
      \item Note. --- Columns for $\Omega_m$, $\langle\Omega_m\rangle$ and $\sigma(\Omega_m)$ refers to the best-estimates of the matter density, the average value and the scatter with respect to $\langle\Omega_m\rangle$. Columns of $w$ have the same meaning. The joint results using the DR measurements of all six redshifts are labeled as "6z". We also show the joint results using DR of the first there redshifts $z_{1,2,3}$ for comparison, labeled as "f3z". While the results obtained from the last three redshifts $z_{4,5,6}$ are labeled as "l3z", and the results by directly multiplying the likelihoods of "f3z" and "l3z" are labeled as "fl3z". Columns of "Magb" refers to the method of calibrating the magnification bias. "\emph{Base.}" represents our baseline analysis, where the magnification bias is calibrated using a fixed cosmology ($\Omega_m=0.315,w=-1$). "\emph{Vary.}" represents the cosmology-dependent calibration, which is shown for consistency check. 
    \end{tablenotes}
\end{minipage}
\end{table*}

\subsubsection{Combined with the BAO data}
Furthermore, we make a comparison between the cosmology implied by DR to that from Dark Energy Spectroscopic Instrument (DESI) DR1 BAO measurements \citep{2024arXiv240403002D} to see if there is any inconsistency. The DESI survey provides robust measurements of BAO feature 
in seven redshift bins from over 6.4 million extragalactic objects in the redshift range $0.1<z<4.2$\citep{2023AJ....165..253H,2023AJ....165...58Z,2023ApJ...944..107C,2023AJ....165..126R,2024arXiv240403002D,2024arXiv240403000D,2024arXiv240403001D,2024AJ....167...62D,2025DESI-DR1}. It represents the current most precise measurements from a galaxy survey. We also present the DE constraints from the previous releases of SDSS BAO measurements for comparison, for which the total galaxy sample size is about 2.8 million and  includes observations of galaxy and quasar samples 
at $z<2.2$ and Ly$\alpha$ forest observations over $2<z<3.5$ \citep{2020MNRAS.499..210N,2021MNRAS.500.1201H,2021MNRAS.501.5616D,2020MNRAS.499.5527T,2021MNRAS.500.3254R,2020MNRAS.498.2492G,2021MNRAS.500..736B}. For redshifts below $z=0.6$, the DESI DR1 data currently covers a smaller effective
volume than SDSS. 

We give the best-fit values, as well as the marginalized mean values, along with the 68\% probability intervals in Table.\ref{Table3}. The 2D-contours of $\Omega_m$ and $w$ for DR in combination with diﬀerent BAO datasets are shown in Fig.\ref{fig:6z-syswt-wcdmcontour-BAO}. We find that the constraints from DR alone are in good agreement with the DESI BAO implied cosmology. The estimated $\Omega_m$ is completely unaffected by including DR compared to DESI BAO alone.  For the DE equation of state parameter $w$, the best-fit value slightly changes from -0.981 to -1.019, with 18\% smaller error bars. However, the conclusion changes when the DESI BAO data is replaced with that from SDSS, where the BAO contours are more orthogonal to DR.
We find that the  best-fit value of $w$ from DR+SDSS BAO changes from -0.69 to -0.871 compared to SDSS BAO alone. Moreover, the inclusion of DR has significantly improved over SDSS BAO constraints of $\Omega_m$ and $w$ by a factor of 2.1 and 1.3, respectively\footnote{We use the MCMC chains and likelihoods from the https://www.sdss.org/science/cosmology-results-from-eboss.}. 
The more significant improvement of the constraint in the $\Omega_m-w$ plane by incorporating DR over SDSS BAO, compared to DESI BAO, can be attributed to three factors: i) the relative locations of the DR contour to that from BAO within the parameter space; ii) the relative degeneracy directions of DR compared to BAO;  iii) the relative sizes of the contour area of BAO and DR. 

In terms of the degeneracy direction of the SDSS BAO data, we discussed this in \cite{2022ApJ...938...72D}, where, for a flat universe, BAO constrains $\Omega_m$ and $w$ through its impacts on $H(z)$. Thus, the degeneracy direction is given by $c\sim-(\partial H/\partial \Omega_m)/(\partial H/\partial w)_{z_\star}$, where ${z_\star}$ being the effective redshift. In the case of a cosmology with $\Omega_m=0.3$ and $w=-1$, $c_{\rm BAO}<0$ while $c_{\rm DR}>0$. Nevertheless, unlike the SDSS BAO data, the DESI contours in Fig.\ref{fig:6z-syswt-wcdmcontour-BAO} are close to vertically aligned. This may be attributed to the redshift distributions of the galaxy samples, as the DESI DR1 galaxies are, on average, from higher redshifts where the sensitivity of $H(z)$ to $w$ is weaker (peaking at $\sim0.6$ and decreasing).

\subsubsection{Combined with the SNe data}
We also compare the constraint from DR to that from PantheonPlus SNe Ia data. The original 'Pantheon sample' includes the full set of spectroscopically conﬁrmed SNe Ia (1,048 SNe Ia) from PanStarrs \citep{10.1117/12.859188} supplemented by SNe Ia observed at low redshift \citep{1999AJ....117..707R,2006AJ....131..527J,2009ApJ...700.1097H,2010AJ....139..519C,2010AJ....139..120F,2011AJ....142..156S}, the SDSS and SNLS samples, and with HST \citep{2012ApJ...746...85S,2007ApJ...659...98R,2014AJ....148...13R,2014ApJ...783...28G,2018ApJ...853..126R}.  Compared to the Pantheon realease, the PantheonPlus SNe Ia data includes six additional samples \citep{2022ApJ...938..113S} 
and comprises 1550 spectroscopically-confirmed SN Ia in the redshift range $0.001 < z < 2.26$, for which the largest improvement in SN numbers is at low redshift. 

We use the public likelihood from \cite{2022ApJ...938..110B}, incorporating the full statistical and systematic covariance. Parameter constraints from the PantheonPlus SNe Ia and DR are shown in Fig.\ref{fig:6z-syswt-wcdmcontour-SN}, where the $\Omega_m-w$ degeneracy direction of DR is found to be completely orthogonal to that of SNe. It is clear from the figure is that the inclusion of DR can strongly affect the parameter constraints compared to SNe alone. Error bars of $\Omega_m$ and $w$ are reduced by 1.73 and 1.5, respectively. Interestingly, we find that the inclusion of DR alleviates the slight tension between $w$ constrained from DESI(SDSS) BAO/SNe. The estimate of $w$ derived from BAO+DR shows better consistency with that from SNe+DR compared to the cases of BAO only and SNe only. Finally, it is noteworthy that DESI BAO, DR and SNe favor $w=-1$ within $1\sigma$, whereas SDSS BAO suggests $w<-1$ at the 2$\sigma$ level.


\begin{table*}
\centering
\caption{Marginalized values and 68$\%$ confidence limits of ($\Omega_m$, $w_0$, $w_a$) estimated from ${\rm DR}_{6z}$ when in combination with DESI BAO/PantheonPlus. \label{Table-w0wa}}
\begin{minipage}{180mm}
\centering
\begin{tabular}{cc|cc|cc|cc}
\hline
\hline
Probes & Magb &$\Omega_m$ (best),  & $\langle\Omega_m\rangle$,  &$w_0$ (best), & $\langle w_0 \rangle$ & $w_a$ (best) & $\langle w_a \rangle$  \\
\hline
        DESI BAO   &-           &$0.361^{+0.027}_{-0.042}$ &$0.344\pm0.037$  &$-0.351_{-0.401}^{+0.201}$ &$-0.559\pm0.330$&$<-1.29$ &$<-1.32$ \\
        BAO+DR   &\emph{Base.}&$0.352^{+0.030}_{-0.045}$ &$0.336\pm0.039$  &$-0.351_{-0.428}^{+0.201}$ &$-0.573\pm0.342$&$<-1.26$ &$<-1.36$\\
\hline
        PantheonPlus   &-           &$0.382_{-0.113}^{-0.051}$ &$0.323\pm0.088$  &$-0.900_{-0.161}^{+0.134}$ &$-0.926\pm0.151$&$0.445_{-1.722}^{+0.334}$& $-0.621\pm1.036$ \\
        DR+PantheonPlus   &\emph{Base.}&$0.343^{+0.042}_{-0.042}$ &$0.346\pm0.044$  &$-0.940_{-0.134}^{+0.107}$ &$-0.957\pm0.128$&$-0.224_{-0.970}^{+0.569}$& $-0.597\pm0.806$\\

\hline
\hline
\end{tabular}
\begin{tablenotes}
      \small
      \item Note. ---Similar to Table.\ref{Table3}, but for the results of a time-varying equation of state parameterized
by $(w_0, w_a)$ model.
    \end{tablenotes}
\end{minipage}
\end{table*}

\begin{figure*}
    \centering
     \subfigure{
     \includegraphics[width=0.8\linewidth, clip]{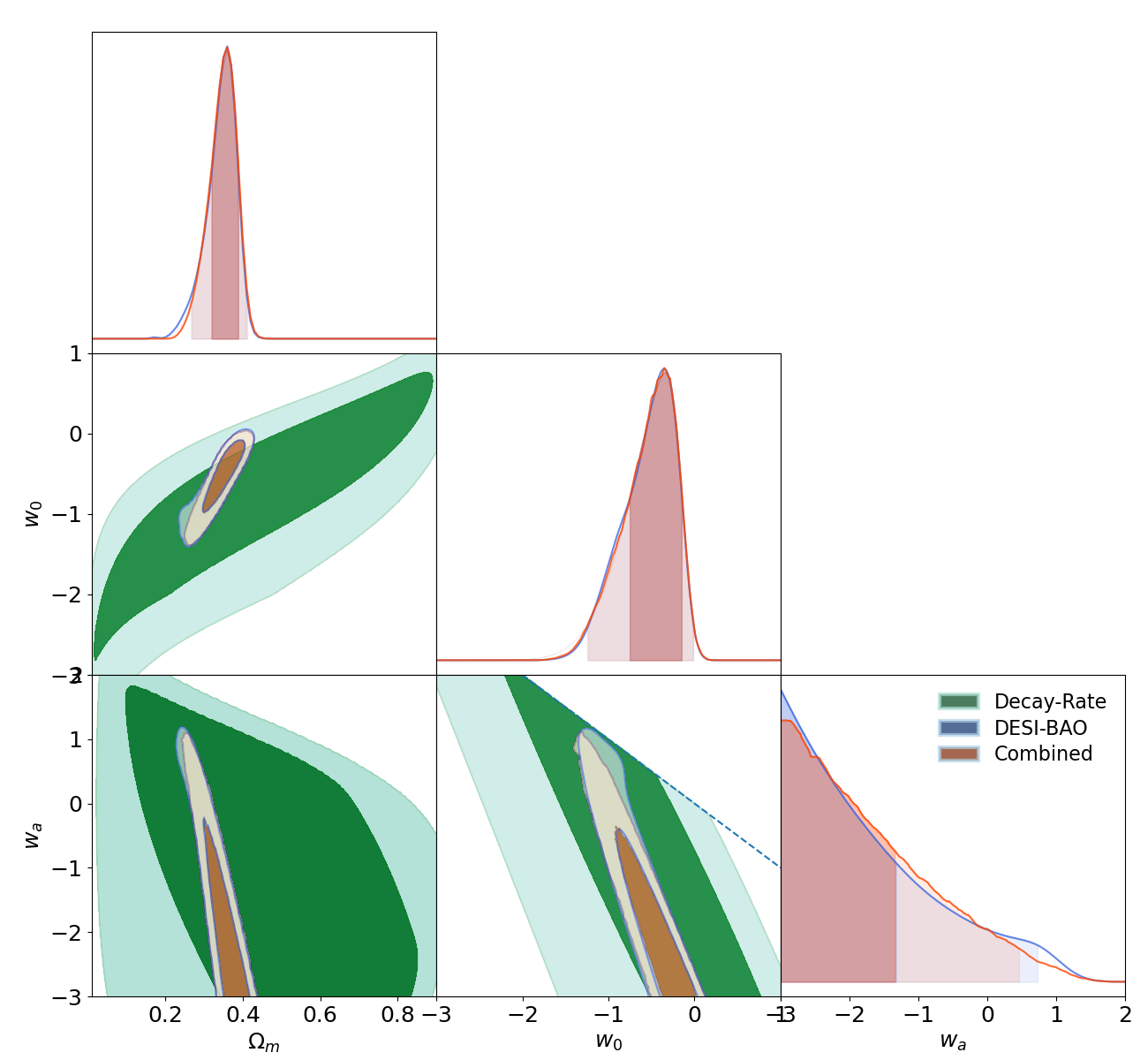}}
   \caption{Constraints on the flat $w_0w_a$ model, from DESI BAO/DR/DESI BAO+DR. The DR measurements align well with the DESI BAO data across all parameter planes. 
   }
\label{fig:6z-syswt-w0wa-BAO}
\end{figure*}

\begin{figure*}
    \centering
     \subfigure{
     \includegraphics[width=0.8\linewidth, clip]{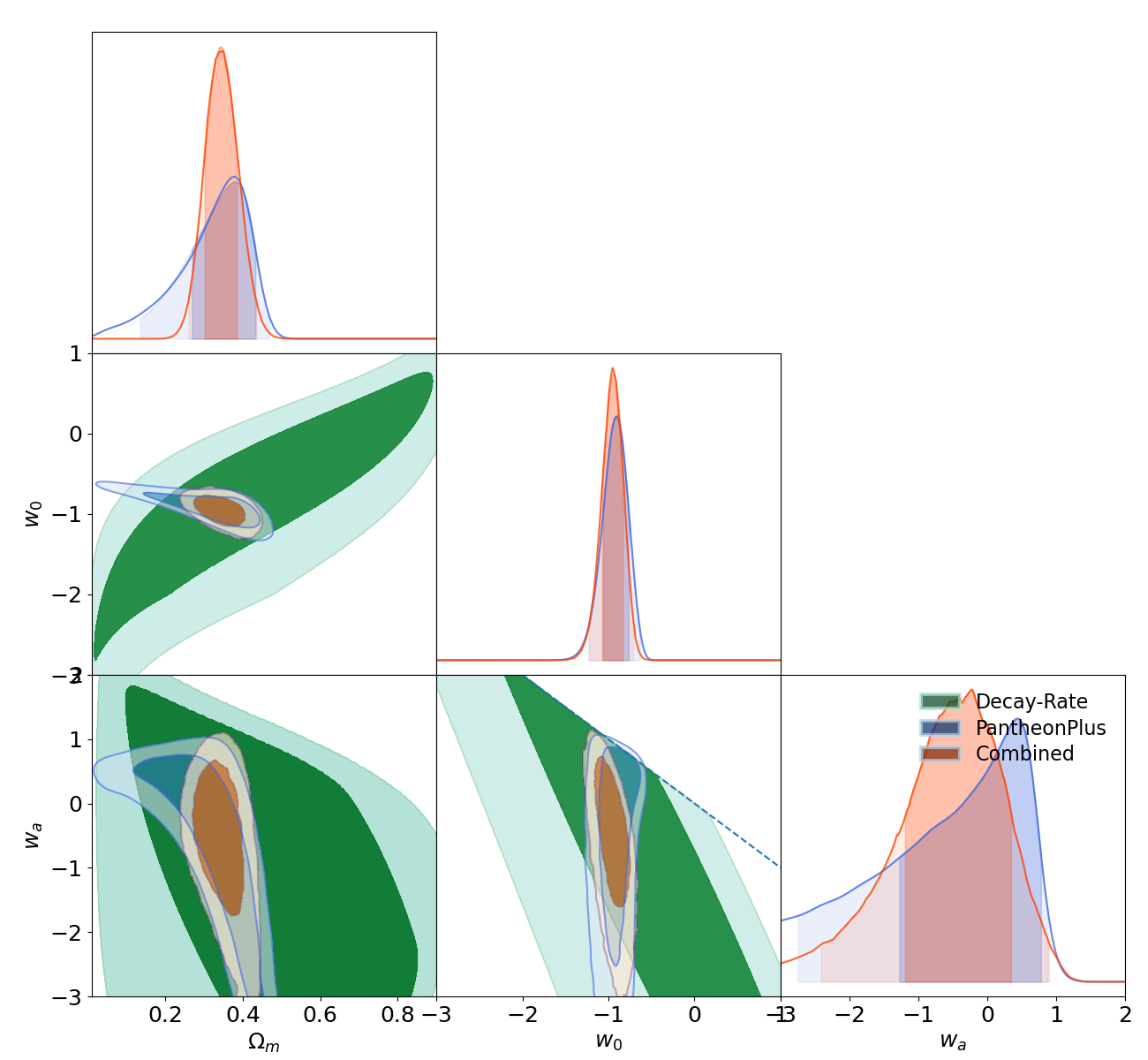}}
   \caption{Similar to Fig.\ref{fig:6z-syswt-w0wa-BAO}, but with constraints from SNe/DR/SNe+DR. }
\label{fig:6z-syswt-w0wa-SN}
\end{figure*}

\subsection{Constraint on a Flat $w_0w_a$ Model}
In this section we discuss the constraining power of DR on a time-varying equation of state, for which the parametrizations of $w$ involves two free parameters\citep{2001IJMPD..10..213C,2003PhRvL..90i1301L}:
\begin{equation}
w(a) = w_0 + (1-a)w_a,    
\end{equation}
where $a=1/(1+z)$ is the scale factor, $w_0$ and $w_a$ both constants. Therefore, DR relies on three cosmological parameters: $\theta = \{\Omega_m, w_0, w_a\}$. It reduces to $\Lambda$CDM model for $w_0=-1$ and $w_a=0$. 

As the BAO feature provides a cosmological probe sensitive to the cosmic expansion history, it is a key science goal of DESI to constrain the $w_0–w_a$ parameter space as well as distinguishing it from $\Lambda$. In this time evolving model, the DESI DR1 BAO data prefers $w_0>-1$ and $w_a<0$, showing a mild pull away from the $\Lambda$CDM value(Table.\ref{Table-w0wa}). 
Therefore, here we analyze the BAO data in conjunction with DR to see if the constraining of $w_0-w_a$ can be further distinguished from $\Lambda$ \footnote{We use the Astropy Package\citep{2019JOSS....4.1298Z} and the Colosseus Package\citep{2018ApJS..239...35D}  to compute the terms of distance and growth factor in order to obtain the theoretical forecast of DR.}. Following \cite{2024arXiv240403002D},  we adopt flat prior ranges for $\Omega_m$, $w_0$ and $w_a$ in the ranges $[0,0.9]$, $[-3,1]$ and $[-3,2]$ respectively, together with the condition $w_0+w_a<0$ imposed to enforce a period of high-redshift matter domination.

Comparisons of the results from DR with those from DESI BAO are shown in Fig.\ref{fig:6z-syswt-w0wa-BAO}, where we present the 2-dimensional marginalized constraints, and Table.\ref{Table3}. The degeneracy directions shown by DR exibit similarities to those of BAO in all planes. The tightest constraints from DR are observed in the $w_0-w_a$ plane, where $w_0$ and $w_a$ are inversely correlated. This anti-correlation arises because both a less negative $w_0$ and a less negative $w_a$ result in a less negative $w(z)$, leading to a longer period of accelerated expansion dominated by DE.
Furthermore, it is evident that $w_0$ is better constrained by BAO/DR than $w_a$, as the latter is weighted by a factor of $z/(1 + z)$ in the expression for $w(z)$, which diminishes the impact of $w_a$. Since the 1$\sigma$ regions of DESI BAO are completely contained within those of DR in the $w_0-w_a$ plane, neither BAO alone nor BAO+DR possesses sufficient power to break the degeneracy between $w_0$ and $w_a$. The constraints from DR in the other two planes seems much weaker. 

Additionally, i) the marginalized posterior constraints from DESI BAO in the $w_0-w_a$ plane show a different degeneracy direction ($c_{\rm BAO>0}$) to that from the $w$CDM model (Fig.\ref{fig:6z-syswt-w0wa-BAO}); ii) the degeneracy direction of $\Omega_m-w_a$ shown by DR is opposite to that of $\Omega_m-w_0$, which is unexpected. These phenomena might be related to the prior ranges chosen for the parameters. Moreover, as the sensitivity of DR/BAO to $w_a$ is weaker, their constraints on $w_a$ are more susceptible by noise in the DR/BAO measurements, as well as the choice of priors.
  
We present the cosmological constraints from SNe for the flat $w_0-w_a$ model in Fig.\ref{fig:6z-syswt-w0wa-BAO}. The results indicate no statistical preference for the $w_0w_a$CDM model. Unlike the case with DESI BAO, the inclusion of DR significantly improves the model constraints.  First, incorporating DR helps to break the degeneracy in the $\Omega_m-w_0$ and $w_0-w_a$ planes. The improvement in the precision of $\Omega_m$ is a factor of 2 over SNe alone, while the error bars for $w_0$ and $w_a$ are reduced by 1.2 and 1.3, respectively. Additionally, we find that the best estimate of $w_a$ shifts by nearly $1\sigma$ with the inclusion of DR, changing from $0.445_{-1.722}^{+0.334}$ to $-0.224_{-0.970}^{+0.569}$. These results are consistent with Fig.\ref{fig:6z-syswt-wcdmcontour-SN}, showing no preference for either $w_0w_a$CDM or $w$CDM over $\Lambda$CDM.

\section{Summary and Discussion}
\label{sec:conclusion}
We report the second model-dependent measurements of gravitational potential decay rate (DR) from observation. These results are based on the DESI DR9 galaxy samples in the redshift range of $0.2\le z<1.4$ and within DECaLS+DES survey areas. These data include a total number of $1.4\times10^7$ galaxies, with the sky coverage being 11183 ${\rm deg^2}$. We detect DR at a total signiﬁcance of $3.1\sigma$. In comparison with using DR at $0.2\le z<0.8$ alone, we find that the including the DR measurements from $0.8\le z<1.4$ is useful to improve its constraining power on the DE model. 

This direct measurement methodology allowed us to harness the potential of the DR to effectively constrain cosmological parameters when combined with other probes. In the $w$CDM model, the DR constraints on DE are consistent well with those from DESI BAO, with the matter density patameter determined to be $\Omega_m=0.292_{-0.014}^{+0.014}$ and the equation-of-state parameter determined to be $w=-1.019_{-0.112}^{+0.105}$ when combining the two probes. The inclusion of DR shows a modest improvement over DESI BAO alone. While the addition of DR shows a significant improvement over SDSS BAO or SNe alone, taking into account the more orthogonal degeneracy directions between them and the slight different locations of the contours. Additionally, all three probes—DR, DESI BAO, and SNe—favor $w=-1$, except for SDSS BAO, which favors $w<-1$. We also explore the constraints of DR on a time-varying equation-of-state model parameterized by $w_0$ and $w_a$. In this case, DESI BAO data favors solutions with $w_0>-1$ and $w_a<0$, and the inclusion of DR does not affect this result. While SNe data prefers $w_0=-1$ and $w_a=0$, and the inclusion of DR not only strengthen this conclusion but significantly improves the precision of parameters. DR+SNe gives $w_0=-0.94_{-0.134}^{+0.107}$, $w_a=-0.224_{-0.970}^{+0.569}$.

\begin{figure*}
    \centering
     \subfigure{
     \includegraphics[width=1\linewidth, clip]{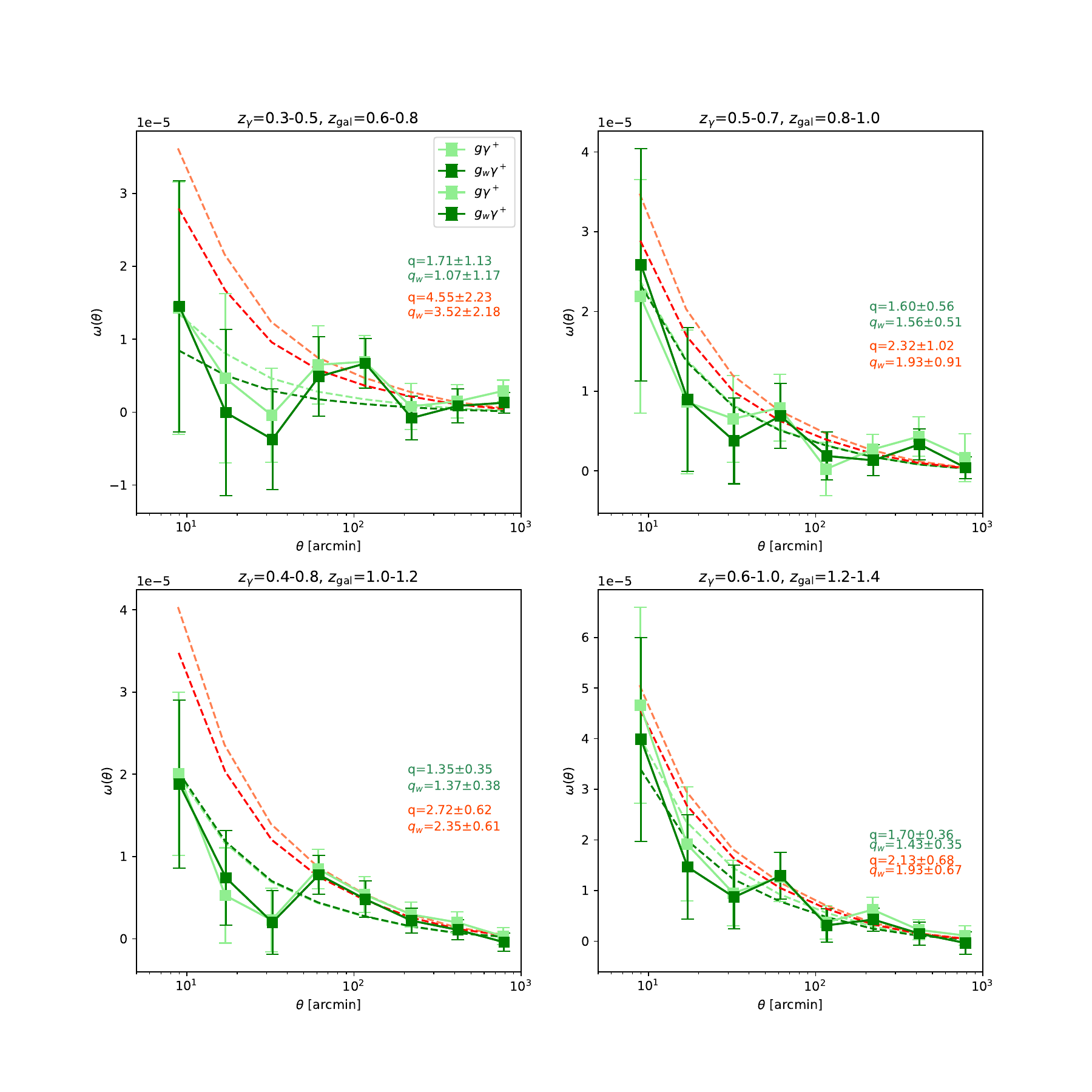}}
   \caption{The measurement of $q$ through the galaxy-shear cross-correlation $\langle\delta_g^L(z_{\rm gal})\gamma^+(z_{gal})\rangle$, where $z_f<z_{\rm gal}$ and $z_{\rm gal}$ the source distribution of galaxies. The values of $q$ are fitted using all the data points (green line) and only the outer data points (red line), where $q_w$ corresponds to the case that that incorporates the imaging weight. In most cases, the value of $q$ is less than three.}
    \label{fig:4q}
\end{figure*}

\begin{figure}
    \centering
     \subfigure{
     \includegraphics[width=1\linewidth, clip]{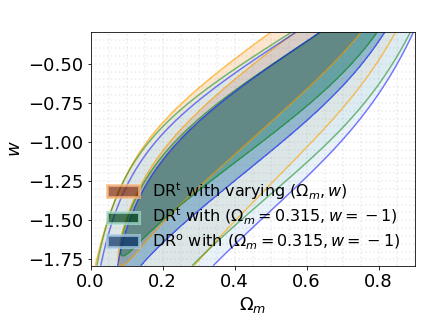}}
   \caption{Constraints on the flat $w$CDM from DR by adopting different strategies for calibrating the magnification bias. The red and green color contours show the results for calibrating the impact of magnification bias on the theoretical prediction of $DR^t$. While the blue color contour is our baseline result, for which the calibration of  magnification bias is conducted on $DR^o$ (DR measurement from observation). For both the green and blue contours, a fixed cosmology is adopted for calibrating the magnification bias. On the other hand, a ($\Omega_m$, $w$)  dependent calibration is adopted for the red color contour. All three constraints show good consistency with each other.}
    \label{fig:magb-omw}
\end{figure}

In the above, we conduct a joint analysis to isolate DR using the cross-power spectrum of $C_{Ig}$ and $C_\phi g$ measured from all six redshift slices. We calibrate the cross-powers for mitigating systematics in the DR measurements, including the imaging systematics and magnification bias. Wherein, to calibrate the influence of magnification bias, we have adopted the Planck cosmology and $q=3$ for estimating its impact. Here, we carry out two more tests for consistency check:

i) First is the measurement of $q$. In our baseline analysis, we derive the value of $q$ from the cumulative number counts of galaxy samples. Here, we conduct a test by measuring $q$ through galaxy-shear cross-correlation. The galaxy density $\delta_g^L(z_i)$ is altered by the magniﬁcation bias due to the mass from $z_f<z_i$. So the magnification bias can be isolated through cross correlating $\delta_g^L(z_i)$ with shear of galaxies from lower redshift: $\langle\delta_g^L(z_i)\gamma^+(z_f)\rangle=q\langle\kappa(z_f<z_i)\gamma^+(z_f)\rangle$.  For simplicity, we theoretically compute the term of $\langle\kappa\gamma^+\rangle^{\rm theory}$ on the right hand of the equation, and derive $q$ by linearly fitting $\langle\delta_g^L(z_i)^{\rm obs}$ with $\langle\kappa\gamma^+\rangle^{\rm theory}$. The results are shown in Fig.\ref{fig:4q}. We find that in most cases, especially for higher redshifts where the magnification bias is larger, the value of $q$ is safely below 3, for which the impact of magnification bias on the DR measurement should be $\lesssim 20\%$ for $\ell\gtrsim10$.

ii) Second is the way to calibrate the magnification bias. The DE constraints in this work are obtained by comparing the theoretical prediction for $DR^t(\Omega_m,w)$ to the $DR^o$ measurement from observation. To be more specific, as presented in Sec.\ref{sec:systematic}, our baseline result (blue color contour in Fig.\ref{fig:magb-omw}) is obtained by calibrating the influence of magnification bias on the cross-correlation measurements of galaxy-ISW and galaxy-lensing from observation for deriving $DR^o$. So for consistency check, instead of conducting the calibration on $DR^o$, here we directly calibrate the theoretical prediction of $DR^t$. The result is shown in the green color contour, for which the agreement with our baseline result is excellent. In the above, a fixed cosmology ($\Omega_m$=0.315, $w=-1$) is adopted for mitigating the magnification bias. Moreover, we also conduct a ($\Omega_m,w$) dependent calibration for deriving $DR^t(\Omega_m,w)$, for which the result is shown as the red color contour. Almost no changes are found in the DE constraint for $\Omega_m<0.5$, demonstrating the rationality of our approach.

Measuring the DE equation of state $w$ is a key science goal of our DR measurement. Besides the $w$CDM model, we also test the DR constraint on a time varying $w(a)$ model, where $w(a)=w_0+(1-a)w_a$. However, the accelerated expansion of the universe can be explained by either DE or modified gravity. Therefore, a detailed test of the modified gravity from DR, as well as the improved measurements of DR by using spectroscopic galaxy samples, will be presented in a series of forthcoming papers.

\section*{Acknowledgments}
FD is supported by the Young Scientists Fund of the National Natural Science Foundation of China (No.12303003). This work was supported by the National Key R\&D Program
of China (2023YFA1607800,\newline
2023YFA1607801, 2020YFC2201602), the China Manned Space Project (\#CMS-CSST-2021-A02), and the Fundamental Research Funds for the Central Universities.
HX is supported by National Science Foundation of China (No.12403010).
The authors thank the HEALPix/healpy software package \citep{Zonca2019,2005ApJ...622..759G}.
The DESI Legacy Imaging Surveys consist of three individual and complementary projects: the Dark Energy Camera Legacy Survey (DECaLS), the Beijing-Arizona Sky Survey (BASS), and the Mayall z-band Legacy Survey (MzLS). DECaLS, BASS and MzLS together include data obtained, respectively, at the Blanco telescope, Cerro Tololo Inter-American Observatory, NSF’s NOIRLab; the Bok telescope, Steward Observatory, University of Arizona; and the Mayall telescope, Kitt Peak National Observatory, NOIRLab. NOIRLab is operated by the Association of Universities for Research in Astronomy (AURA) under a cooperative agreement with the National Science Foundation. Pipeline processing and analyses of the data were supported by NOIRLab and the Lawrence Berkeley National Laboratory (LBNL). Legacy Surveys also uses data products from the Near-Earth Object Wide-field Infrared Survey Explorer (NEOWISE), a project of the Jet Propulsion Laboratory/California Institute of Technology, funded by the National Aeronautics and Space Administration. Legacy Surveys was supported by: the Director, Office of Science, Office of High Energy Physics of the U.S. Department of Energy; the National Energy Research Scientific Computing Center, a DOE Office of Science User Facility; the U.S. National Science Foundation, Division of Astronomical Sciences; the National Astronomical Observatories of China, the Chinese Academy of Sciences and the Chinese National Natural Science Foundation. LBNL is managed by the Regents of the University of California under contract to the U.S. Department of Energy. The complete acknowledgments can be found at https://www.legacysurvey.org/acknowledgment/. The Photometric Redshifts for the Legacy Surveys (PRLS) catalog used in this paper was produced thanks to funding from the U.S. Department of Energy Office of Science, Office of High Energy Physics via grant DE-SC0007914.

\appendix
\section{Cross-correlations}
We calculate these two lensing angular power spectra $C_\ell^{g\kappa}$, $C_\ell^{\mu\kappa}$ using the Limber approximation\citep{1953ApJ...117..134L}:

\begin{equation}
C_{\ell}^{g\kappa}(z_f,z_b)=b_f\int^\infty_0 dz\frac{W_g(z,z_f)g(z,z_b)}{\chi^2}P(k,z),  \\
\end{equation}
where $P(k,z)$ is the 3D mattter power spectrum for a wave number $k\simeq(\ell+1/2)/\chi(z)$ and at a redshift z, $W_g$ the normalized galaxy window function and $g(z,z_b)$ the lensing weight function, defined as
\begin{equation}
g(z,z_0)=\left(\frac{3H_0^2}{2c^2}\Omega_{m,0}\right) (1+z)\chi(z)\int^\infty_zdz^\prime\frac{\chi(z^\prime)-\chi(z)}{\chi(z^\prime)}W(z^\prime,z_0).    
\end{equation}

\begin{eqnarray}
C_{\ell}^{\mu\kappa}(z_f,z_b)=q_f \int^\infty_0 &dz&\frac{c}{H(z)}\frac{W_\mu(z,z_f)g(z,z_b)}{\chi^2}P(k,z), \nonumber \\
\end{eqnarray}
where
\begin{equation}
W_\mu(z)=\left(\frac{3H_0^2}{2c^2}\Omega_{m,0}\right)(1+z)g(z,z_f).
\end{equation}

Similarly, these two ISW angular power spectra $C_\ell^{gI}$, $C_\ell^{\mu I}$ can be simplified to 

\begin{equation}
C^{gI}_\ell=\frac{4}{(2\ell+1)^2}\int dzW_{\rm{ISW}}(z)W_g(z,z_f)b_fP(k,z),
\end{equation}

where  D(z) is the linear growth factor and $W_{\rm{ISW}}$(z) the ISW window function, defined as
\begin{equation}
W_{\rm{ISW}}(z)\equiv 3\Omega_m T_0\left(\frac{H_0}{c}\right)^2\frac{d}{dz}\left[\frac{D(z)}{a(z)}\right].
\end{equation}

\begin{equation}
C^{\mu I}_\ell=\frac{4}{(2\ell+1)^2}\int^z_0 dzW_{\rm{ISW}}(z)W_\mu(z)P(k,z)q.
\end{equation}

\bibliography{dr}

\end{document}